\documentstyle[12pt,cite]{article}
\renewcommand{\thefootnote}{\fnsymbol{footnote}}
\begin{document}

\begin{titlepage}

{\hfill DFPD/92/TH/46}

{\hfill IC-MATH/9-92}

{\hfill hep-th/9310066}

\vspace{0.4cm}

\centerline{{\large \bf LIOUVILLE EQUATION AND SCHOTTKY
PROBLEM}\footnote[5]{Partly
supported by a SERC fellowship and by
the European Community Research Programme `Gauge Theories,
applied supersymmetry and quantum gravity', contract SC1-CT92-0789}}

\vspace{0.8cm}

\centerline{\large{\sc Marco} {\sc Matone}\footnote{e-mail:
matone@padova.infn.it, mvxpd5::matone}}

\vspace{0.4cm}

\centerline{\it Department of Mathematics}
\centerline{\it Imperial College}
\centerline{\it 180 Queen's Gate, London SW7 2BZ, U.K.}

\vspace{0.1cm}

\centerline{\it and}

\vspace{0.1cm}

\centerline{\it Department of Physics ``G. Galilei'' - Istituto
Nazionale di Fisica Nucleare} \centerline{\it University of Padova}
\centerline{\it Via Marzolo, 8 - 35131 Padova,
Italy\footnote[7]{Present address}}

\vspace{1.2cm}

\centerline{\large ABSTRACT}

\vspace{0.6cm}

An Ansatz for the Poincar\'e metric on compact Riemann surfaces is
proposed. This implies that the Liouville equation
reduces to an equation resembling a non chiral analogous
of the higher genus relationships (KP equation) arising in the framework of
Schottky's problem solution. This approach connects uniformization
(Fuchsian groups) and moduli space theories with KP hierarchy.
Besides its mathematical interest, the Ansatz has some applications in the
framework of quantum Riemann surfaces arising in 2D gravity.

\vspace{0.2cm}

\end{titlepage}\newpage
\setcounter{footnote}{0}\renewcommand{\thefootnote}{\arabic{footnote}}

\renewcommand{\theequation}{\thesection.\arabic{equation}}
\newcommand{\mysection}[1]{\setcounter{equation}{0}\section{#1}}

\mysection{Schottky Problem And KP Hierarchy}

 Let us consider a genus $h$ compact Riemann surface
$\Sigma$. A fundamental object defining the complex structure
of $\Sigma$ is the Riemann period matrix
\begin{equation}
\Omega_{ij}\equiv\oint_{\beta_i}\omega_j,
\label{aa1}\end{equation}
where the $\omega_k$'s denote the $h$ holomorphic differentials with the
standard normalization $\oint_{\alpha_i}\omega_j=\delta_{ij}$. By means
of the Riemann bilinear relations it can be proved that
$\Omega_{ij}$ is symmetric and has positive definite imaginary part
(see for example \cite{fk1}).
Let us consider the Siegel space
\begin{equation}
{\cal A}_h={\cal H}_h/Sp(h,{\bf Z}),
\label{aa2}\end{equation}
where ${\cal H}_h$ denotes the Siegel upper-half plane, that is the
space of symmetric $h\times h$ matrices with
positive definite imaginary part.
To recognize  the locus in ${\cal A}_h$ of the Riemann
period matrices is the famous Schottky problem. This problem has been
 solved essentially by Dubrovin, Mulase and Shiota
\cite{dubrovin,mulase,shiota}. The solution is based on the proof of the
Novikov conjecture stating that
\begin{equation}
u(x,y,t)=2\partial_x^2 \log \Theta (Ux+Vy+Wt+z_0|\Omega),
\label{S1}\end{equation}
satisfies the KP equation if and only if
$\Omega$ is the period matrix of some $\Sigma$.
The corresponding equations
on $\Omega$ (see eq.(\ref{Boris})) were derived in \cite{dubrovin} where it
was proved that they determine an algebraic variety
with a component given by the matrices of the $\beta$-periods.
In \cite{shiota} Shiota pointed out that if $u$
in eq.(\ref{S1}) satisfies the KP equation,
then there are vectors $U^k$, such that the function
\begin{equation}
u(t_1,t_2,\ldots)=2\partial_t^2\log \Theta
\left(\sum_{k=1}^\infty U^kt_k|\Omega\right),
\qquad t_1=x,\, t_2=y,\, t_3=t,
\label{S2}\end{equation}
determines solutions of the KP hierarchy
\begin{equation}
\left[{\partial\over
 \partial t_j}-L_j,{\partial\over \partial t_k}-L_k\right]
=0,\label{S3}\end{equation}
where the order $k$ differential operators $L_k$
have coefficients depending on $\vec{t}\equiv (t_1,t_2,\ldots)$
  and are determined by the equation $(\partial_{t_k}-L_k)
\psi(\vec{t},z)=0$, $\psi$ being the
Baker-Akhiezer function on $\Sigma$.
Since the space of vectors $U^k$ is $h$-dimensional,
 there are two commuting operators of coprime order
which are linear combinations of the $L_k$'s.
Therefore one can apply the results in \cite{krichever1}
to show that $\Omega$ is the Riemann matrix of the surface
defined by these operators.

\mysection{The Ansatz}

 Let $\Sigma$ be a compact Riemann surface of genus $h>1$.
It is well-known that the Liouville equation on $\Sigma$
\begin{equation}\partial_z\partial_{\bar z}\varphi(z,\bar z)={1\over 2}e^{
\varphi(z,\bar z)},\label{1l}\end{equation}
is uniquely satisfied by the Poincar\'e metric (with Gaussian curvature
$-1$).
This metric can be written in terms of the
inverse map of uniformization
\begin{equation}
J^{-1}_H:\Sigma\longrightarrow H,
\label{e1}\end{equation}
where $H=\{w|{\rm
Im} \, w > 0\}$ denotes the upper half plane.
The Poincar\'e metric on $H$ is
\begin{equation}
ds^2={|dw|^2\over({\rm Im} \, w)^2},
\label{e2}\end{equation}
so that on $\Sigma\cong H/\Gamma$ (here $\Gamma$ is a hyperbolic
Fuchsian group)
\begin{equation}
e^{ \varphi(z,\bar {z})}={|{J_H^{-1}(z)}'|^2\over({\rm Im}\,
J_H^{-1}(z))^2},\label{e3}\end{equation}
which is invariant under $SL(2,{\bf R})$ fractional transformations of
$J^{-1}_H$.
Unfortunately no one has succeeded in writing down
$J^{-1}_H$ in terms of the moduli of $\Sigma$.

Here we consider
the following Ansatz for the Poincar\'e metric\footnote{Notice
that a possible choice for the matrix to be positive definite is to set $A_{ij}
={\Omega_{ij}^{(2)}}^{-1}$, in this case (\ref{prova11}) coincides
with the Bergman metric.}
\begin{equation}e^\varphi=\sum_{i,j=1}^h\omega_iA_{ij}\overline\omega_j.
\label{prova11}\end{equation}
To get the inverse map one has to solve the Schwarzian equation
\begin{equation}\{J_H^{-1},z\}=
T^F(z),\qquad \label{4}\end{equation}
where
\begin{equation}
T^F(z)=\varphi_{zz}-{1\over 2}\varphi_z^2,
\label{stress1}\end{equation}
is the classical Liouville stress tensor
(or Fuchsian projective connection).
By (\ref{prova11}) we have
\begin{equation}
T^F(z)={2\sum_{i,j=1}^h\omega_i''A_{ij}\overline \omega_j-
3\left(\sum_{i,j=1}^h\omega_i'A_{ij}
\overline\omega_j\right)^2\over 2\left(\sum_{i,j=1}^h\omega_iA_{ij}
\overline\omega_j\right)^2}.
\label{stress2}\end{equation}
Observe that eq.(\ref{1l}) implies that
\begin{equation}
\partial_{\bar z}T^F(z)=0.
\label{stress2er}\end{equation}

Eq.(\ref{4}) can be reduced to the linear equation
\begin{equation}
\left(2\partial^2_z+T(z)\right)\psi=0.
\label{ficxk}\end{equation}
Actually it turns out that, up to $SL(2,{\bf C})$ linear fractional
transformations,
\begin{equation}
J^{-1}=\psi_1/\psi_2,\label{g1000}\end{equation}
with $\psi_1$ and $\psi_2$ two linearly independent solutions
of (\ref{ficxk}) (see \cite{mma} for a discussion on this point).

Inserting (\ref{prova11}) in (\ref{1l}), the
Liouville equation becomes\begin{equation}{\sum_{i,j,k,l=1}^h\omega_l^2
\partial_z\left({\omega_{i}/\omega_l}\right)A_{ij}A_{lk}\overline\omega_k^2
\partial_{\bar z}\left({\overline\omega_j/\overline\omega_k}\right)}=\left(
\sum_{i,j=1}^h\omega_iA_{ij}\overline\omega_j\right)^3.\label{2l}\end{equation}

This equation has a strict similarity with the relations between the
periods of holomorphic differentials on Riemann surfaces \cite{dubrovin}.
Thus one should expect that $A_{ij}$ depends on the moduli
through the Riemann period matrix.
To show this similarity, we write down the fundamental relations
given in \cite{dubrovin}. Let us introduce the following notation
$$
U_k=-\omega_k(P),
$$
$$
V_k=-\omega'_k(P),
$$
\begin{equation}
W_k=-{1\over 2}\omega''_k(P)-{1\over 2}c(P)U_k,
\label{w2}\end{equation}
where $c(P)$ is a projective connection \cite{dubrovin} and $P$ is an
arbitrary point on $\Sigma$.
In \cite{dubrovin} Dubrovin proved that the function (\ref{S1})
is a solution of the KP equation
\begin{equation}
u_{yy}=(4u_t-6uu_x-u_{xxx})_x,
\label{dgpl}\end{equation}
if and only if the following relations between $U,V,W,\Omega$ and
an additional constant $d$ are satisfied
(see \cite{dubrovin} for
notation)
\begin{equation}
\sum_{i,j,k,l=1}^hU_iU_jU_kU_l\widehat \Theta_{ijkl}[n]+
\sum_{i,j=1}^h\left({3\over 4}V_iV_j-U_iW_j\right)
\widehat \Theta_{ij}[n]+d \widehat \Theta [n]=0, \qquad n\in{\bf Z}^h_2.
\label{Boris}\end{equation}
We emphasize that this result is a fundamental step
to solve Schottky's problem.

Our remark is that eq.(\ref{2l}) looks like a non chiral generalization
of (\ref{Boris}). In the notation introduced above eq.(\ref{2l}) reads
\begin{equation}\sum_{i,j,k,l=1}^h
\left(U_lV_i-U_iV_l\right)A_{ij}A_{lk}
\left(\overline U_k\overline V_j-\overline U_j \overline V_k\right)
=\left(
\sum_{i,j=1}^hU_iA_{ij}\overline U_j\right)^3.
\label{2lh}\end{equation}
We stress that solving this equation is equivalent to solving crucial
questions arising in uniformization theory, Fuchsian groups and related
subjects. In particular, Weil-Petersson's 2-form $\omega_{WP}$
can be recovered using
the fact that the classical Liouville action evaluated at the classical
solution is the K\"ahler potential of $\omega_{WP}$ \cite{0}.

Another aspect that should be investigate
is whether eq.(\ref{2lh}) furnishes conditions on the period
matrix in a more manageable form than KP equation (\ref{dgpl})-(\ref{Boris}).

A possible approach to study eq.(\ref{2l}) is using
Krichever-Novikov's differentials $\psi_j^{(n)}$ \cite{kn}.
These differentials are holomorphic on $\Sigma\backslash\{P_+,P_-\}$
with prescribed behaviour at $P_\pm$.
In particular, in terms of local coordinates
$z_\pm$ vanishing at $P_\pm\in \Sigma$, we have
\begin{equation}
\psi_j^{(n)}(z_\pm)(dz_\pm)^n=a_j^{(n)\pm}
z_\pm^{\pm j -s(n)}\left(1+{\cal
O}(z_\pm)\right)\left(dz_\pm\right)^n,
\quad s(n)={h\over 2}-n(h-1),\label{onehalf}\end{equation}
where $j\in{\bf Z}+h/2$ and $n\in{\bf Z}$.
By the Riemann-Roch theorem, $\psi_j^{(n)}$ is uniquely fixed
by choosing the value of $a_j^{(n)+}$ or $a_j^{(n)-}$. In the
following we set $a_j^{(n)+}=1$.

These differentials can be written in terms of theta
functions\footnote{In the appendix we illustrate the method to construct
differentials in higher genus Riemann surfaces.} \cite{cmp}
\begin{equation}
\psi_j^{(n)}(z)=C_j^{(n)}
\Theta\left(I(z)+{\cal D}^{j;n}|\Omega\right)
 {\sigma(z)^{2n-1} E(z, P_+)^{j-s(n)}\over
 E(z,P_-)^{j+s(n)}},
\label{psij}\end{equation}
where
$$
{\cal D}^{j;n}=\left(j-s(n)\right)I(P_+)-
\left(j+s(n)\right)I(P_-)+(1-2n)\Delta,
$$
and constant $C_j^{(n)}$ is fixed by
the condition $a_j^{(n)+}=1$.

Let  ${\cal C}$ be a homologically trivial
contour separating $P_+$ and $P_-$.
The dual of $\psi_j^{(n)}$  is defined by
\begin{equation}
{1\over 2\pi i}\oint_{\cal C} \psi_j^{(n)}
\psi^k_{(n)}=\delta_j^k,\label{dualpsi}\end{equation}
which implies
\begin{equation}
\psi^j_{(n)}=\psi_{-j}^{(1-n)}.\label{easy1}\end{equation}
Note that (\ref{onehalf}) provides a basis for the
$1-2s(n)=(2n-1)(h-1)$ holomorphic $n$-differentials
on $\Sigma$ ($h\ge 2$)
\begin{equation}
{\cal H}^{(n)}=\left\{\psi_k^{(n)}\big |s(n)\le k\le -s(n)\right\},
\qquad n\ge 2.
\label{qdrtcdfab}\end{equation}
Furthermore, from
\begin{equation}
\widetilde{\cal H}^{(m)}=\left\{\psi_k^{(m)}
\big |1-s(m)\le k\le s(m)-1\right\},\qquad m\le -1,
\label{qdrtcdfabc}\end{equation}
one can define the space of generalized Beltrami differentials.
They are vanishing everywhere on $\Sigma$ except in a disk
where coincide with \cite{mma}
\begin{equation}
\widetilde{\cal B}^{(m)}=\left\{\partial_{\bar z}\psi_k^{(m)}
\big |1-s(m)\le k\le s(m)-1\right\},\qquad m\le -1,
\label{qdrtcdfabcdd}\end{equation}
 (for $m=-1$ one gets the Beltrami differentials considered in
\cite{cmp}). Observe that the differentials in
(\ref{qdrtcdfabc}) have poles both in $P_+$ and $P_-$.
In particular, $\widetilde{\cal H}^{(1-n)}$ is the dual space of
${\cal H}^{(n)}$.

We now  expand the holomorphic 3-differentials in
(\ref{2l}) in terms of the basis introduced above. We have
\begin{equation}
\omega_i^2\partial_z(\omega_j/\omega_i)=\sum_{p=1}^{5h-5}a_{ij}^p\psi_{p+s(3)-1
}^{(3)},\qquad a_{ij}^p={1\over 2\pi i}\oint_{{\cal C}}\psi_{-p-s(3)+1}^{(-2)}
\omega_i^2\partial_z(\omega_j/\omega_i),\label{i1}\end{equation}\begin{equation}
\omega_i\omega_j\omega_k=\sum_{p=1}^{5h-5}b_{ijk}^p\psi_{p+s(3)-1}^{(3)},\qquad
b_{ijk}^p={1\over2\pi i}\oint_{{\cal C}}\psi_{-p-s(3)+1}^{(-2)}\omega_i\omega_j
\omega_k.\label{i2}\end{equation}
Inserting these expansions in (\ref{2l}) we
get the `Liouville relations'
\begin{equation}\sum_{i,j,k,l=1}^ha_{ij}^pA_{ik}A_{jl}\overline a^q_{kl}=
\sum_{i,j,k,l,m,n=1}^hb_{ijm}^pA_{ik}A_{jl}A_{mn}\overline b_{kln}^q.\label{i3}
\end{equation}\\Let us notice that the coefficients $a^q_{kl}$ and $b_{kln}^q$
are functionals of the holomorphic differentials and their derivatives computed
at $P_+$ and coincide with the vectors of $\beta$-periods of second-kind
differentials.

The above expansions provide relations involving the holomorphic differentials,
theta functions and their derivatives. To see this it is
sufficient to notice that the coefficients $a^p_{ij}$ and
$b^p_{ijk}$ are vanishing for $p<1$ and $p>5h-5$. The reason is that
in this range the $\psi_{-p-s(3)+1}^{(-2)}$'s
are holomorphic in $P_-$ or $P_+$.
This implies that for $p<1$ and $p>5h-5$, the contribution to $a^p_{ij}$ and
$b^p_{ijk}$ coming from the poles at $P_-$ or $P_+$
add to zero. Notice that this `residue formula' is crucial to
get important relations such as Hirota's formulation of the KP
hierarchy (see for example \cite{johnfay}).

\mysection{The Accessory Parameters}

 Here we consider some aspects concerning the Fuchsian accessory
parameters. First of all we introduce the projective connection
 \begin{equation}
T^S(z)=\left\{J^{-1}_\Omega,z\right\},
\label{proOmega}\end{equation}
where $J_\Omega:\Omega\to\Sigma$ denote the Schottkian uniformization map.
Here $\Omega$ denotes the region of discontinuity in
$\widehat {\bf C}={\bf C}\cup\{\infty\}$
of the Schottky group ${\cal S}$ and $\Sigma \cong\Omega/{\cal S}$.
Let us introduce the following notation for the Krichever-Novikov vector
fields and quadratic differentials
\begin{equation}
e_k\equiv \psi_k^{(-1)},\qquad\quad
\Omega^k\equiv \psi_{-k}^{(2)}.
\label{fdgs}\end{equation}
Let ${\cal T}_\Sigma$ be the holomorphic projective connection on
$\Sigma$ obtained from the symmetric bidifferential of the second-kind with
bi-residue 1 and zero $\alpha$-periods.
The Fuchsian accessory parameters $\lambda_1^{(F)},...,
\lambda_{3h-3}^{(F)}$ and the Schottkian accessory parameters
$\lambda_1^{(S)},...,\lambda_{3h-3}^{(S)}$ are defined by
\begin{equation}
T^F={\cal T}_\Sigma+\sum_{k=1}^{3h-3}
\lambda_k^{(F)}\Omega^{k+1-h_0}, \qquad
T^S={\cal T}_\Sigma+\sum_{k=1}^{3h-3}
\lambda_k^{(S)}\Omega^{k+1-h_0},\quad h_0\equiv
{3\over 2}h. \label{m2}\end{equation}

In order to write ${\cal T}_\Sigma$ explicitly we consider
 an arbitrary nonsingular point $f$ of the theta divisor,
that is $\Theta(f)=0$ and grad$\,\Theta(f)\ne 0$.
 We define
\begin{equation}
H_f(z)=\sum_{k=1}^h\Theta_{k}(f)\omega_k(z),
\label{m34}\end{equation}
\begin{equation}
Q_f(z)=\sum_{j,k=1}^h\Theta_{jk}(f)\omega_j(z)
\omega_k(z),\label{m35}\end{equation}
\begin{equation}
T_f(z)=\sum_{i,j,k=1}^h\Theta_{ijk}(f)
\omega_i(z)\omega_j(z)\omega_k(z).\label{m36}\end{equation}
The holomorphic projective connection is \cite{fay}
\begin{equation}
{\cal T}_\Sigma(z)=\left\{\int^z_{P_0} H_f,z\right\}+
 {3\over 2} \left(Q_f(z)\over H_f(z)\right)^2-2{T_f(z)\over H_f(z)}.
 \label{m37}\end{equation}
At a zero of $H_f$ we have
\begin{equation}
Q_f(z_0)=\pm H_f'(z_0),
\qquad T_f(z_0)=-H_f''(z_0)\pm {3\over 2}Q_f'(z_0),
\label{m38}\end{equation}
with the sign $\pm$ chosen accordingly as
 $\Theta(z-z_0\mp f)\equiv 0,\; \forall z\in\Sigma$.

Besides $T^F$ and $T^S$, also ${\cal T}_\Sigma$ can be expressed
as a Schwarzian derivative. To do this we simply note that
according to the general rule described above the equation
\begin{equation}
\left({\partial^2\over \partial z^2}+{1\over2}{\cal T}_\Sigma(z)
\right)\phi(z)=0,
\label{new1aqw}\end{equation}
has as solutions two linearly independent
$-{1\over 2}$-differentials $\phi_1,\,\phi_2$, satisfying
the equation
\begin{equation}
{\cal T}_\Sigma(z)=\left\{{\phi_2/ \phi_1},z\right\}.
\label{dfsfd}\end{equation}

Note that the Fuchsian accessory parameters are given by
\begin{equation}
\lambda_k^{(F)}={1\over 2\pi i}\oint_{{\cal C}}\left(
 \left\{J_H^{-1}(z),z\right\}
-\left\{J_\Sigma^{-1}(z),z\right\}
\right)e^{k+1-h_0},
\label{m39}\end{equation}
where
\begin{equation}
J_\Sigma^{-1}(z)={\phi_2/ \phi_1}.
\label{m391}\end{equation}
It is interesting to note that the integrand resembles
the chain rule for the Schwarzian derivative
\begin{equation}
\{w(t(z)),z\}(dz)^2-\{t(z),z\}(dz)^2=
\{w(t),t\}(dt)^2,
\label{oiioj}\end{equation}
in particular
\begin{equation}
\left\{J_H^{-1}\left(J_\Sigma^{-1}(z)\right),z\right\}
-\left\{J_\Sigma^{-1}(z),z\right\}=
\left\{J_H^{-1}\left(J_\Sigma^{-1}\right),
J_\Sigma^{-1}\right\}\left(\partial_zJ_\Sigma^{-1}(z)\right)^2.
\label{oiiojb}\end{equation}

We stress that the accessory parameters can be written
as a line integral of a one-form written in terms of theta
functions and holomorphic differentials.
In particular for the Fuchsian accessory parameters we have
$$
\lambda_k^{(F)}={1\over 2\pi i}\oint_{{\cal C}}\left(
{2\sum_{i,j=1}^h\omega_i''A_{ij}\overline \omega_j-
3\left(\sum_{i,j=1}^h\omega_i'A_{ij}
\overline\omega_j\right)^2\over 2\left(\sum_{i,j=1}^h\omega_iA_{ij}
\overline\omega_j\right)^2}
-\left\{\int^z_{P_0} H_f,z\right\} \right.
$$
\begin{equation}
\left. -{3\over 2} \left(Q_f(z)\over H_f(z)\right)^2
+2{T_f(z)\over H_f(z)}\right)e^{k+1-h_0}.\label{maa39}\end{equation}

In the second reference in \cite{0},
where the results for the punctured Riemann sphere are generalized
to higher genus Riemann surfaces, a relationship has been established between
$c_k^{(h)}=\lambda_k^{(F)}-\lambda_k^{(S)}$, the Liouville action
evaluated on the classical solution
and the Weil-Petersson metric.
In particular it turns out that
\begin{equation}
{1\over 2}{\partial S^{(h)}_{cl}\over \partial z_i}=c_i^{(h)},
\qquad {\partial c_i^{(h)}\over \partial \bar z_j}=-{1\over 2}
\left \langle
 {\partial\over \partial z_i}\,,{\partial\over \partial z_j
}\right \rangle_{WP},
\label{addendum1}\end{equation}
where the brackets denote the Weil-Petersson metric on the
Teichm\"uller space $T_h$
projected onto the Schottky space whose coordinate are
 $z_1,...,z_{3h-3}$.
Since the difference
\begin{equation}
\Theta(z)=T^F(z)-T^S(z)=
\sum_{k=1}^{3h-3}c_k^{(h)}\Omega^{k+1-h_0}(z),\label{difference}\end{equation}
 is a holomorphic quadratic differential
(i.e. a section of $T^\star T_h$),
the formulas in eq.(\ref{addendum1}) are equivalent to
 \begin{equation}
\partial S^{(h)}_{cl}=2\Theta,
\qquad {\overline\partial}\partial S^{(h)}_{cl}=-2i\omega_{WP},
\label{41}\end{equation}
where $d=\partial+\overline\partial$ is the exterior differentiation
on the Schottky space and $\omega_{WP}$ is the Weil-Petersson 2-form
on this space. Because the Schottky projective connection
depends holomorphically on the moduli we have
\begin{equation}
\overline\partial
T^F= - i\omega_{WP},\label{wppc}\end{equation}
that by (\ref{stress2}) gives
\begin{equation}
\omega_{WP}=i\overline
\partial{2\sum_{i,j=1}^h\omega_i''A_{ij}\overline \omega_j-
3\left(\sum_{i,j=1}^h\omega_i'A_{ij}
\overline\omega_j\right)^2\over 2\left(\sum_{i,j=1}^h\omega_iA_{ij}
\overline\omega_j\right)^2}.
\label{stress2ba}\end{equation}

Similar results have been derived by Fay \cite{fay2}. In particular it
turns out that
\begin{equation}
\left\{J_H^{-1},z\right\}= {\cal T}_\Sigma-24\pi i
\sum_{j,k=1}^h \left( {\partial\over \partial \Omega_{jk}}
\log c_0\right) \omega_j(z)\omega_k(z),\label{fay11}\end{equation}
where
\begin{equation}
c_0=\left[{8\pi^2 {\rm det}' \Delta\over {\rm det}\, {\rm
Im}\,\Omega}\right]^{-1/2},\label{anomalia}\end{equation}
is the anomaly in the spin-1/2 bosonization formula computed
with respect to the Poincar\'e metric $e^\varphi$.

The connection with the Weil-Petersson metric
on $T_h$  arises if we consider the quasi-conformal mapping
\begin{equation}
\partial_{\bar z} f^\rho =\rho \partial_z f^\rho,
\qquad \rho=t_1\nu_1+t_2\nu_2.\label{quasiabbastanza}\end{equation}
It turns out that
\begin{equation}
 -24\pi \partial\overline\partial
\log c_0 =\big<\nu_1,\nu_2\big>_{WP},\label{fay12}\end{equation}
where $\big<\nu_1,\nu_2\big>_{WP}=\int_\Sigma e^\varphi
\nu_1\overline\nu_2$ and
\begin{equation}
\partial=\partial_{t(p)}=\sum_{j,k=1}^h {\partial\over \partial \Omega_{jk}}
\delta \Omega_{jk},\label{schiffer1}\end{equation}
is the Schiffer variation (see \cite{fay2} for details).

Another possible way to investigate eq.(\ref{prova11}) is
by noticing that both the first and second variations vanish for
the deformation of the
complex structure induced by the harmonic Beltrami differentials
\cite{ahlfors,wolpert3}.
Applying this condition to (\ref{prova11}) should give further informations on
the form of matrix $A_{ij}$.

As a final remark, we observe that, besides any mathematical interest, the
solution for the Poincar\'e metric is crucial to get
explicit expressions for correlators in string theory. In particular in
the `uniformization approach' to 2D quantum gravity
\cite{tak1,mma} one needs the explicit expression for the Liouville action
evaluated at the classical solution to compute the `VEV of quantum Riemann
surfaces' $\langle\Sigma\rangle$ (see \cite{tak2,mmb}).

\appendix

\mysection{Appendix}

Let us introduce the theta function with characteristic
\begin{equation}
\Theta \left[^a_b\right]\left(z|\Omega\right)=
\sum_{k\in {\bf Z}^h}e^{\pi i (k+a)\cdot\Omega\cdot(k+a)+
2\pi i  (k+a)\cdot (z+b)},
\qquad\Theta\left(z|\Omega\right)\equiv\Theta
\left[^0_0\right]\left(z|\Omega\right),
\label{oixho}\end{equation}
where $z\in{\bf C}^h,\; a,b\in{\bf R}^h$.
When $a_k,b_k\in\{0,1/2\}$,  $\Theta \left[^a_b\right]
\left(z|\Omega\right)$  is even or odd
depending on the parity of $4a\cdot b$. The $\Theta$-function
is multivalued under a lattice shift in the $z$-variable
\begin{equation}
\Theta \left[^a_b\right]\left(z+n+\Omega\cdot m|\Omega\right)=
e^{-\pi i m \cdot\Omega\cdot m -2\pi i m \cdot z
+2\pi i (a\cdot n-b\cdot m)}
\Theta\left[^a_b\right]\left(z|\Omega\right).
\label{ojcslc}\end{equation}
An important object to construct differentials in higher
genus is the prime form
$E(z,w)$. It is a holomorphic
$-1/2$-differential both in $z$ and $w$,
vanishing for $z=w$ only
\begin{equation}
E(z,w)={\Theta\left[^a_b\right]\left(I(z)-I(w)|\Omega\right)\over
h(z)h(w)}. \label{pojdlk}\end{equation}
Here $h(z)$ denotes
the square root of $\sum_{k=1}^h\omega_k(z)
\partial_{u_k}\Theta\left[^a_b\right]\left(u|\Omega\right)|_{u_k=0}$;
it is the holomorphic 1/2-differential with non singular
(i.e. $\partial_{u_k}\Theta\left[^a_b\right]
\left(u|\Omega\right)|_{u_k=0}\ne 0$) odd spin structure
$\left[^a_b\right]$.
The function $I(z)$ in (\ref{pojdlk})
denotes the Jacobi map
\begin{equation}
I_k(z)=\int_{P_0}^z\omega_k,\qquad z\in\Sigma,
\label{ghftdt}\end{equation}
with $P_0\in \Sigma$ an arbitrary base point.
This map is an embedding of $\Sigma$ into
the Jacobian
\begin{equation}
J(\Sigma)={\bf C}^h/L_\Omega,\qquad L_\Omega={\bf Z}^h
+\Omega {\bf Z}^h.\label{jac}\end{equation}
By (\ref{ojcslc}) it follows that the multivaluedness of $E(z,w)$ is
\begin{equation}
E(z+{n}\cdot {\alpha} +{m}\cdot {\beta},z)=
e^{-\pi i m\cdot \Omega\cdot m -2\pi i m\cdot
\left(I(z)-I(w)\right)}E(z,w).
\label{primeform}\end{equation}
In terms of $E(z,w)$ one can construct the following
$h/2$-differential with empty divisor
\begin{equation}
\sigma(z)=\exp\left(-\sum_{k=1}^h\oint_{\alpha_k}\omega_k(w)\log
E(z,w)\right),\label{sigma}\end{equation}
whose multivaluedness is
\begin{equation}
\sigma(z+{n}\cdot {\alpha} +{m}\cdot {\beta})=
e^{\pi i (h-1) {m} \cdot \Omega\cdot {m}-2\pi i
{m}\cdot \left(\Delta-(h-1){I}(z)\right)}
\sigma(z),\label{mltvld4}\end{equation}
where $\Delta$ is (essentially) the {\it vector of Riemann constants}
\cite{fay}.
Finally we quote two theorems:
\begin{itemize}
\item[{\bf a.}]{{\bf Abel Theorem} \cite{fk1}. {\it A necessary
and sufficient condition for ${\cal D}$ to be the divisor
of a meromorphic function is that}
\begin{equation}
I\left({\cal D}\right)=0\; {\rm mod}\, \left(L_\Omega\right)\;
and \; {\rm deg}\, {\cal D}=0.\label{abella}\end{equation}}
\item[{\bf b.}]{{\bf Riemann vanishing theorem} \cite{fay}.
{\it The function}
\begin{equation}
\Theta\left(I(z)-\sum_{k=1}^hI(P_k)+\Delta\bigg|\Omega\right),
\qquad z,P_k\in \Sigma,\label{rvth}\end{equation}
{\it either vanishes identically or else it has
$h$ zeroes at $z=P_1,\ldots,P_h$}.}
\end{itemize}

We are now ready to explicitly construct the differential
$f^{(n)}$ defined above. First of all note that
\begin{equation}
\widetilde f^{(n)}=\sigma(z)^{2n-1}
{\prod_{k=h+1}^p E(z, P_k)\over
\prod_{j=1}^{p-2n(h-1)} E(z, Q_j)},\label{explkpr}\end{equation}
is a multivalued $n$-differential
with ${\rm Div}\, \widetilde f^{(n)}=
\sum_{k=h+1}^pP_k-\sum_{k=1}^{p-2n(h-1)}Q_k$. Therefore we set
\begin{equation}
f^{(n)}(z)=g(z)
\widetilde f^{(n)},\label{explk}\end{equation}
where, up to a multiplicative constant, $g$ is fixed
by the requirement that $f^{(n)}$ be
singlevalued. From the multivaluedness of
the $E(z,w)$ and $\sigma(z)$
it follows that, up to a multiplicative constant
\begin{equation}
g(z)=\Theta\left(I(z)+{\cal D}
\big|\Omega\right),
\label{thetafncte}\end{equation}
with
\begin{equation}
{\cal D}=\sum_{k=h+1}^pI(P_k)-
\sum_{k=1}^{p-2n(h-1)}I(Q_k)+(1-2n)\Delta.
\label{iudlkm}\end{equation}
By Riemann vanishing theorem $g(z)$ has just $h$-zeroes
$P_1,\ldots,P_h$ fixed by  ${\cal D}$. Thus the
requirement of singlevaluedness also fixes the position of the
remainder $h$ zeroes.
To make manifest the divisor in the RHS of
(\ref{explk}) we first recall that the
image of the canonical line bundle $K$ on the Jacobian
of $\Sigma$ coincides
with $2\Delta$ \cite{fay}.
 On the other hand, since
\begin{equation}
\left[K^n\right]=
\left[ \sum_{k=1}^pP_k-\sum_{k=1}^{p-2n(h-1)}Q_k\right],
\label{abel}\end{equation}
by Abel theorem we have\footnote{The square brackets in (\ref{abel})
denote the divisor class associated to the line bundle $K^n$.
Two divisors belong to the same class if they differ by a divisor
of a meromorphic function.}
\begin{equation}
{\rm Div}\, \Theta\left(I(z)+{\cal D}
\big|\Omega\right)={\rm Div}\, \Theta\left(I(z)-\sum_{k=1}^hI(P_k)+
\Delta\bigg|\Omega\right),
\label{thetafncte1}\end{equation}
and by Riemann vanishing theorem
\begin{equation}
{\rm Div}\, \Theta\left(I(z)+{\cal D}
\big|\Omega\right)=\sum_{k=1}^hI(P_k).
\label{thetafncte2}\end{equation}


\begin{thebibliography}{99}
\bibitem{fk1} H.M. Farkas and I. Kra, {\it Riemann Surfaces},
II ed. Springer-Verlag 1992.
\bibitem{dubrovin} B.A. Dubrovin, Russian Math. Surv. {\bf 36} (1981) 11;
Math. USSR-Izv. {\bf 19} (1982) 285.
\bibitem{mulase} M. Mulase, J. Diff. Geom. {\bf 19} (1984) 403.
\bibitem{shiota} T. Shiota, Inv. Math. {\bf 83} (1986) 333.
\bibitem{krichever1} I.M. Krichever, Funktz.
 Analiz. Prilozh. {\bf 11} n.1 (1977) 15.
\bibitem{0} P.G. Zograf and L.A. Takhtajan,
Math. USSR Sbornik, {\bf 60} (1988) 143; {\bf 60} (1988) 297.\\ L.A. Takhtajan,
Proc. Symp.  Pure Math. Amer. Math. Soc.
{\bf 49}, part 1 (1989) 581.
\bibitem{mma} M. Matone, {\it Uniformization Theory And 2D Gravity. I.
Liouville Action And Intersection Numbers}, preprint IC-Math/8-92,
DFPD/92/TH/41, hepth/9306150.
\bibitem{kn} I.M. Krichever and S.P. Novikov, Funktz.
 Analiz. Prilozh. {\bf 21} n.2 (1987) 46;
  n.4 (1987) 47; {\bf 23} n.1 (1989) 24.
\bibitem{cmp} L. Bonora, A. Lugo, M. Matone and
J. Russo, Comm. Math. Phys. {\bf 123} (1989) 329.
\bibitem{johnfay} J.D. Fay, Proc. Symp. in Pure Math. {\bf 49} (1989) 485.
\bibitem{fay} J.D. Fay,
{\it Theta Functions On Riemann Surfaces},
Lect. Notes in Math. {\bf 356} (1973).
\bibitem{fay2} J.D. Fay,
{\it Perturbation Of
Analytic Torsion On Riemann Surfaces}, Memoirs of the AMS, {\bf 96},
n.464 (1992).
\bibitem{ahlfors} L.V. Ahlfors, Ann. Math. {\bf 74} (1961) 171;
J. Anal. Math. {\bf 9} (1961) 161.
\bibitem{wolpert3} S.A. Wolpert, Invent. Math. {\bf 85} (1986) 119.
\bibitem{tak1} L.A. Takhtajan, in `New Symmetry Principles In
Quantum Field Theory', Eds. J. Fr$\ddot{\rm o}$hlich et al., Plenum Press,
1992.
\bibitem{tak2} L.A. Takhtajan, {\it Liouville Theory: Quantum
Geometry Of Riemann Surfaces}, preprint hepth/9308125.
\bibitem{mmb} M. Matone, {\it Quantum Riemann Surfaces, 2D Gravity And The
Geometrical Origin Of Minimal Models}, preprint DFPD/93/TH/62,
hepth/9309096.
\end{thebibliography}
\end{document}